\documentclass{article}

\usepackage[final,nonatbib]{neurips_2023}

\usepackage{xcolor}
\usepackage{graphicx}
\usepackage{algorithm}
\usepackage{algpseudocode}

\usepackage{bm}
\usepackage{bbm}
\usepackage{amsmath}
\usepackage{amssymb}
\usepackage{amsthm}
\usepackage{multirow}
\usepackage{subcaption}
\usepackage{comment}

\theoremstyle{definition}
\newtheorem{definition}{Definition}

\newtheorem{property}{Property}

\newcommand{\func}[1]{\textbf{\textcolor{teal}{#1}}}  
\newcommand{\baseC}[1]{\bf{\textcolor{violet}{#1}}}  
\newcommand{\ttime}[1]{\textbf{\textcolor{brown}{#1}}}  

\title{Onchain Sports Betting using UBET Automated Market Maker}

\author{%
  Daniel Jiwoong Im, Alexander Kondratskiy, Vincent Harvey, Hsuan-Wei Fu \\
  UBET Sports\\
  ubetsports.io\\
  \texttt{general@ubetsports.io} \\
}

\begin{document}

\maketitle

\begin{abstract}
The paper underscores how decentralization in sports betting addresses drawbacks of traditional centralized platforms, ensuring transparency, security, and lower fees. Non-custodial solutions empower bettors with ownership of funds, bypassing geographical restrictions. Decentralized platforms enhance security, privacy, and democratic decision-making. However, decentralized sports betting necessitates automated market makers (AMMs) for efficient liquidity provision. Existing AMMs like Uniswap lack alignment with fair odds, creating risks for liquidity providers. To mitigate this, the paper introduces UBET AMM (UAMM), utilizing smart contracts and algorithms to price sports odds fairly. It establishes an on-chain betting framework, detailing market creation, UAMM application, collateral liquidity pools, and experiments that exhibit positive outcomes. UAMM enhances decentralized sports betting by ensuring liquidity, decentralized pricing, and global accessibility, promoting trustless and efficient betting.
\end{abstract}

\section{Introduction}
The sports betting industry is experiencing significant growth and shows promising prospects for a bright future. With the increasing popularity of online betting platforms, the industry has witnessed substantial expansion in recent years \cite{BRI2023, Davies2023}. Additionally, the rise of mobile betting apps has made wagering more convenient and accessible to a larger audience. The sports betting market is poised to thrive in the global market as the industry continues to evolve and adapt to changing consumer demands.

Sports betting boasts a long history, originating in ancient civilizations and evolving into a multi-billion-dollar global market \cite{Charpentier2019, Matheson2021}. However, traditional sports betting suffers from several pitfalls. Centralized platforms controlled by bookmakers or agencies may lack transparency and accessibility, while users face issues of trust and potential manipulation of odds. Not to mention that users often take a huge risk  from their custody solutions \cite{KPMG2022, Aquilina2023}. High fees, including transaction costs and commissions on winning bets, can deter bettors. 

Decentralization in sports betting addresses several limitations and challenges present in traditional centralized betting platforms. It ensures transparency and fairness through immutable records of all transactions and bets using blockchain technology \cite{Nakamoto2008}. This trustless betting system eliminates the need for users to place their faith in a central authority, instilling confidence and security among bettors. Lower fees on decentralized platforms, attributed to the removal of intermediaries, make betting more accessible and appealing to a broader audience. 
Unlike traditional sportsbook which maintains bettors' capital, having a non-custodial solution enables bettors to hold their capital in their own wallet \cite{Wood2023}. Moreover, geographical restrictions become a thing of the past, as decentralized platforms can navigate regulatory complexities with greater ease, opening up a global market. Enhanced security and privacy, coupled with near-instant payouts, provide a seamless and reliable betting experience. By incorporating community governance models, users have a say in platform decisions, fostering a more democratic and user-centric environment. 

Decentralized sports betting requires automated market makers (AMMs) to facilitate efficient and continuous liquidity provision for the betting markets. AMMs are essential in the decentralized finance (DeFi) ecosystem and play a crucial role in ensuring the smooth functioning of decentralized sports betting platforms. Unfortunately, the existing AMMs like Uniswap \cite{Adams2020UniswapVC}, Curve \cite{Egorov2021}, etc., are not suitable for pricing sports betting odds due to a lack of alignment with fair odds, which consequently exposes liquidity providers to elevated risks through arbitrage opportunities. To address this issue, we propose a solution that tackles the main drawback by applying UBET AMM (UAMM) \cite{uamm2023} to price sports betting odds. UAMMs use algorithms to determine the prices of bets based on the current supply of funds in the liquidity pool while referencing other sportsbook odds to compute the estimation of the markets' fair odds.

In this paper, we leverage smart contract techniques and UAMM algorithms to establish an on-chain sports betting framework and conduct experiments to analyze sports betting market behaviour. Our main contributions are as follows:
\begin{enumerate}
    \item We introduce a formal methodology for creating sports betting markets on a blockchain. This involves defining, minting, and merging conditional tokens, calculating payouts, and handling reporting and redemption.
    \item We demonstrate a novel application of UAMM specifically for sports betting. UAMM utilizes the swap rate function to determine exchange rates between conditional tokens, providing odds that minimize impermanent loss for liquidity providers.
    \item We propose an additional collateral liquidity pool that allows bettors to use base currency like USDC, streamlining the sports betting process. This approach distills the common features of leading UAMM implementations, and we provide funding calculations for managing conditional tokens and liquidity pool shares.
    \item We conduct extensive controlled and uncontrolled experiments with repeated trials, including sensitivity analysis for various factors. Our results show positive permanent gains while maintaining a relatively low vigorish.
\end{enumerate}
In conclusion, UAMM plays a vital role in decentralized sports betting platforms by ensuring continuous liquidity, decentralized pricing, and global accessibility while eliminating intermediaries, fostering trustless betting, and providing an efficient betting environment for users worldwide.

\section{Related Work}

Sports betting technology has undergone significant advancements over the years, propelled by the emergence of online platforms, mobile apps, and blockchain-based solutions. Within the realm of traditional sports betting, extensive and thorough examination has been conducted to explore factors that influence user behaviour, including user experience, interface design, and the impact of bonuses and promotions in attracting bettors to online betting platforms \cite{Guillou2021}. Furthermore, exhaustive research has been dedicated to investigating the convenience, accessibility, and portability of mobile betting applications \cite{PaySafe2020}.
In contrast, blockchain-based sports betting solutions are relatively new and we provide a brief overview of the most recent approaches to web3 sportsbooks.

{\bf Web3 Sportsbooks} 
Most of the web3 sportsbooks provide a non-custodial wallet solution where bettors connect to the platform using their blockchain wallet and the bet transactions get recorded in the blockchain. However, certain sportsbooks, such as DexSport \cite{DexSport}, still retain central authority to ban bettors and control the payout process despite utilizing blockchain wallets. In contrast, our UBET platform \cite{UBET} is designed to automatically send out payouts to users' wallets for settled conditional tokens.

To the best of our knowledge, all web3 sportsbooks (e.g., SportX \cite{SportX}, Azuro \cite{Azuro}, Divvy \cite{Divvy}, etc.) set their odds off-chain, which does not fully eliminate central authority and results in high vigorish. In contrast, we employ the UAMM algorithm \cite{uamm2023} to determine odds on-chain, ensuring transparency through publicly available smart codes, fairness to all bettors, and lower vigorish."

{\bf Prediction Markets} 
A web3 prediction market is a decentralized platform that facilitates the buying and selling of prediction contracts for future events, including sports events. This creates a trustless and transparent ecosystem for forecasting. Examples of prediction markets like Augur \cite{augur}, HedgeHog Market \cite{hedgehogmarket}, Polymarket \cite{Polymarket}, Manifold Market \cite{manifold}, etc., utilize Uniswap or a variant of Uniswap to incorporate available information into prices. However, despite prediction markets aggregating dispersed information to forecast outcomes which eventually measure the fair price reflecting true probabilities, the initial and convergence process prices may be mispriced. This mispricing provides arbitrage opportunities for bettors, resulting in impermanent loss for liquidity providers. To address this issue, Uniswap \cite{Adams2020UniswapVC} leverages estimated fair prices using external market prices to avoid significant mispricing. This approach reduces arbitrage opportunities as the estimated fair price converges towards the true probability of events. It remains an open question what is the best way to measure or forecast the fair price?

\section{Preliminary - Sports Betting on Blockchain}
Sports betting can be viewed as a form of discrete option contract, wherein the maturity date aligns with the game's end date and time. The act of placing a bet on a particular outcome corresponds to the establishment of terms and conditions, and the actual match result triggers the exercise of these contractual events. To illustrate, a binary option contract could be created for a betting market centred around the question, {\em "Which team will emerge victorious between Chicago Bulls and Toronto Raptors on November 7, 2023?"}. Bettors are afforded the opportunity to engage in contract trading until the game start. The value of these contracts is contingent on the market consensus price which is the probability of the specific event outcome materializing. As the maturity date approaches, the contract price will eventually converge to either one or zero, contingent upon the actual event outcome.

To initiate a betting market, the following factors need to be considered: i. the possible outcomes (conditional tokens), ii.  betting period. iii. resolution mechanism (the oracle), iv. methodical settlement of conditional tokens, and v. payout mechanism. In the subsequent sections, the underlying mechanisms of each aspect are presented.

\begin{figure}[t]
    \centering
    \includegraphics[width=\textwidth]{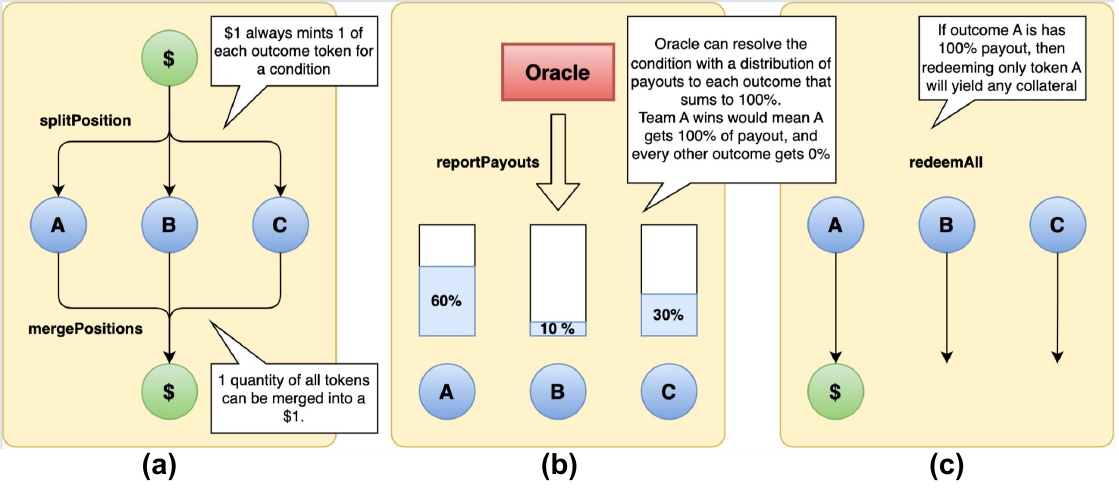}
    \caption{Conditional tokens are a powerful tool used in betting markets to differentiate contracts based on different event outcomes. 
    The conditional tokens allow users to bet on specific conditions or outcomes within a given event, adding depth and granularity to the betting market.
    Here is the depiction of (A) Token minting and merging (B) Outcome resolution (C) Payouts for each outcome}
    \label{fig:cond_tokens}
\end{figure}

\subsection{Conditional Tokens}
Given a betting market, we distinguish contracts for different event outcomes using conditional tokens \cite{gnosis_cond}.
By utilizing these tokens, bettors gain the opportunity to place bets on specific conditions or outcomes within a given event. 
The quantity of conditional tokens corresponds to the magnitude of the payout in the event of a correct outcome.

Formally, let $\mathcal{T} = (1\tau_1, 1\tau_2, \cdots, 1\tau_K)$ consists of each $K$ unique outcome conditional token.
The number of conditional tokens represents the ownership of their bet on one of $K$ finite possible outcomes of events.
The price of each token $p\tau_i$ is determined by the automated market maker (AMM)
and it corresponds to a market-estimated probability of an event outcome. 
As a consequence, the sum of the price of each token adds up to 1, $\sum_i p\tau_i = 1$.
Once the event result is determined, the probability of the correct outcome becomes one and the rest becomes zero.
For example, Manchester United vs. Chelsea has three outcomes, Man United Winning, Draw, and Chelsea Winning.
These outcomes will be represented by the three conditional tokens ($\tau_{\text{MU}}, \tau_{\text{Draw}}, \tau_{\text{Chelsea}}$).
Suppose that the probabilities of the three outcomes are 25\%, 50\%, and 25\% , then
the $p\tau_\text{MU}$, $p\tau_\text{Draw}$, and $p\tau_\text{Chelsea}$ are \$0.25, \$0.5, and \$0.25 respectively.

\subsection{Minting \& Merging Conditional Tokens}
Token minting refers to the process of creating new tokens on a blockchain. 
In the context of conditional tokens, we mint $K$ conditional tokens $\mathcal{T}$ when there is a collateral token that is backed.
For example, if the collateral token is {\em USDC} stablecoin, then we mint $1$ USDC-backed $K$ conditional tokens.
This is possible because the price of $K$ tokens sums up to 1, which is worth $1$ USDC.
We denote collateral token as $\tau_0$ going forward.

The concept of token burning is sought-after in the blockchain. 
The tokens are deliberately destructed from existing tokens, often to reduce the total supply or maintain scarcity. 
For conditional tokens, we use token merging instead to return the collateral.
We return collateral tokens by merging the $K$ minted conditional tokens $\mathcal{T}$ to get back $1\tau_0$.
The $K$ conditional tokens $\mathcal{T}$ are removed from the supply in the merging process.
See the illustration in Figure~\ref{fig:cond_tokens}(a) for the minting and merging process of conditional tokens.

\subsection{Payout Calculation}
\label{sec:payout_calc}
The payout distribution is determined based on the total number of tokens held by each participant for the correct outcome.
The odds are set during the betting period and any fees or commissions are deducted by the platform.
How do we determine the payout amount based on the $K$ minted conditional tokens $\mathcal{T}$?
Suppose a user bets $1\tau_0$ on $i^{th}$ outcome event. 
Here, we describe the procedure of how the payouts (the odds) are determined.
First, we mint K conditional tokens $\mathcal{T}$ by having $1\tau_0$ collateral backed.
Among $\mathcal{T}= (\tau_1, \tau_2, \cdots, \tau_K)$, we keep $\tau_i$ and exchange the rest of $\tau_j$, $j\neq i$ for $\tau_i$ at some exchange rates.
We return $d\tau_i$ conditional tokens to users where
\begin{align*}
    d\tau_i = \tau_i + \sum_{j\neq i} \func{swap}(\tau_j, i; \Gamma),
\end{align*}
$\text{swap}(\tau_j, i; \Gamma)$ swaps $\tau_j$ for some $\tau_j$, and $\Gamma$ is a liquidity pool parameter (see Algorithm~\ref{alg:odd_calc}).
The rest of $\tau_j$ conditional tokens are given to the liquidity provider who took the opposite side of the bet.
The exchange rate is determined by AMM such as Uniswap \cite{Adams2020UniswapVC}, Curve \cite{Egorov2021}, and UBET AMM \cite{uamm2023}. 

\subsection{Reporting Payouts and Redemption}
The resolution of conditional tokens in a betting market refers to the process of determining the outcome of a specific event that bettors have bet on. 
Once the event is resolved, the associated conditional tokens are settled based on the actual outcome. 
The oracle is the only entity capable of resolving a condition. 
Usually, the oracle is a trusted source of external information that reports the result to the blockchain network. They can be decentralized or centralized, depending on the platform's design. The oracle's address is specified when the condition is created and cannot be changed afterward.
Figure~\ref{fig:cond_tokens}(b-c) illustrates what happens once the oracle resolves the market. 
The $A$ conditional token, which is the outcome specified by the oracle, becomes worth $1\tau_0$ each but the rest of the other conditional tokens is worth $0\tau_0$. Bettors who hold $A$ conditional tokens are eligible for a payout, while those who hold tokens associated with incorrect outcomes may lose their bets

\begin{algorithm}[t]
\caption{Odds Calculation}\label{alg:odd_calc}
\begin{algorithmic}
\Require $i\in [1,K]$ is the outcome index to bet on.
\Require $d\tau_0 \geq 0$ is the wager size in base currency $\tau_0$.
\State $odd \leftarrow d\tau_0$
\For{$j=1\cdots K$, $j \neq i$}
    \State // Swap $\tau_j$ tokens for $\tau_i$ tokens 
    \State $odd \leftarrow odd + \func{swap}(d\tau_j, i; \Gamma)$.
\EndFor
\State Return $odd$.
\end{algorithmic}
\end{algorithm}

\section{Applying UBET AMM to Sports Betting}
\label{sec:uamm}
Just as determining a fair price is essential in trading, the core of sports betting lies in the identification of fair odds. Multiple methods can be employed to calculate these odds. A conventional approach involves manual setting and adjustment of odds, often based on the detection of movements by sharp bettors or emulation of reputable sportsbooks. Additionally, sportsbooks utilize statistical models to analyze extensive data, encompassing factors such as team form, head-to-head records, home-field advantage, and player injuries, among others. The primary objective is to predict the probabilities of diverse outcomes in a given event.

In the domain of decentralized exchanges, automated market makers (AMMs) play a crucial role by facilitating trades and determining asset prices, akin to odds in the context of sports betting. These AMMs employ dynamic pricing algorithms based on the demand and supply of assets within the pool. 
The utilization of AMM offers several notable advantages, including operation on blockchain networks, thereby endowing users with complete control over their funds and trades, engendering transparency and fostering trust in the system. Moreover, the constant provision of liquidity ensures the availability of liquid markets for various assets, mitigating the risk of illiquid conditions. Nonetheless, it is important to acknowledge that existing AMMs suffer from a significant drawback: a lack of alignment with fair odds, which consequently exposes liquidity providers to elevated risks through arbitrage opportunities. To address this issue, we present UBET AMM (UAMM) and its application to sports betting in the subsequent section, proposing a solution that tackles the aforementioned limitation of prevailing AMMs.

In the context of sports betting, during the payout calculation process, we apply UAMM for swapping between two or more conditional tokens as mentioned in Section~\ref{sec:payout_calc}. More specifically, the minted conditional tokens not chosen by a bettor, $d\tau_j$ for all $j \neq i$, will be swapped for the conditional tokens $\tau_i$, the outcome selected by a bettor. UAMM determines the exchange rate between the two tokens $\tau_j$ and $\tau_i$.

\subsection{Add \& Remove Funds}
We present Add and Remove functions for UAMM. While the original UAMM's add and remove functions take conditional tokens as inputs and outputs $\func{Add}(d\tau_1,d\tau_2,\cdots,d\tau_K)$ and $\func{Remove}(d\tau_1,d\tau_2,\cdots,d\tau_K)$, we utilize collateral token $\tau_0$ as input and output for these two functions $\func{Add}(d\tau_0)$ and $\func{Remove}(d\tau_0)$. This approach aligns more naturally with betting markets in general, as bettors use collateral tokens for betting, and conditional tokens can be merged into collateral tokens. In other words, we introduce a collateral liquidity pool alongside other conditional token liquidity pools, aiming to minimize holding the conditional tokens by merging them whenever possible and maintaining the collateral tokens instead. 

When LPs add funds to the collateral, they go to the collateral liquidity pool. 
Let $s_{lp}$ be the number of LP tokens that you receive when you add funds either in base currency $\baseC{\tau_0}$ or $K$ minted tokens.
$\func{Add}(d{\tau_0}) \rightarrow s_{lp}$ returns a new-minted amount of LP shares, 
\begin{align*}
    s_{lp}= d{\tau_0} \cdot \frac{TS}{TV}.
\end{align*}
We update all the state of AMM before we calculate $s_{lp}$, that is re-compute the liquidity pool balances ($R\tau_0, R\tau_1, \cdots,  R\tau_K)$, total value $TV$, total investment balance $TB$. Finally, update $TS$. 
[TODO: Check whether the update happens before or after]. 

When LPs remove funds, we return the money from the collateral liquidity pool, along with a portion of conditional tokens, if applicable. 
We calculate how many collateral tokens to take out based on the number of shares $s_{lp}$.
$\func{Remove}(s_{lp}) \rightarrow d{\tau_0}$ takes $s_{lp}$ as input and returns the collateral tokens and some $K$-minted tokens, where 
\begin{align*}
    d\tau_i =  R\tau_0 \cdot \frac{s_{lp}}{TS}
\end{align*}
where $R\tau_0$ is the collateral liquidity pool balance at time $T$.  
While the market is still open, the conditional tokens that are merged get stored in the collateral liquidity pool $R\tau_0$ and the remaining unmerged conditional tokens stay in the liquidity pools $R\tau_i$, $i\neq 0$.
Because unmerged conditional tokens are locked in the liquidity pool, LP can only remove a portion of their funds that are available in the collateral liquidity pool $R\tau_0$.

Here are the following Additivity and Reversibility properties:
\begin{property}[Additivity]
    The Add and Remove transactions are addictive while the fair prices remain the same.
    That is, the result of the states are the same whether a user performs two of the same successive add or remove transactions, or through a single transaction:
    \begin{enumerate}
        \item 
            $\func{Add}(d{\tau_0}) + \func{Add}(d{\tau_0}^\prime) 
                    \Longleftrightarrow \func{Add}(d{\tau_0}+d{\tau_0}^\prime)$
        \item $\func{Remove}(s_{lp}) + \func{Remove}(s_{lp}^\prime) \Longleftrightarrow \func{Remove}(s_{lp}+s_{lp}^\prime)$
    \end{enumerate}
\end{property}

\begin{property}[Reversibility]
    The Add and Remove transactions are reversible while the fair prices remain the same.
    The state derived from Add operation is reversible by Remove, and visa versa:
    \begin{enumerate}
        \item $\func{Add}(\func{Remove}(s_{lp})) \rightarrow s_{lp}$
        \item $\func{Remove}(\func{Add}(d{\tau_0})) \rightarrow (d{\tau_0})$
    \end{enumerate}
\end{property}

The proof of Additivity are the same as one from the original paper and the proof of Remove and Reversibility are almost the same except that we replace $R\tau_i$ to $R\tau_0$ (see the Appendix in \cite{uamm2023}).

\subsection{UAMM: UBET Automated Market Maker}
The fundamental concept behind the UAMM sets it apart from conventional AMMs. Uniswap's reliance on internal market data often leads to significant price disparities compared to other exchanges, particularly when the internal market volume is relatively small compared to the external market. This discrepancy creates arbitrage opportunities for bettors while exposing Liquidity Providers (LPs) to potential losses. Conversely, the UAMM approach adheres to the principles of efficient market theory, wherein we estimate fair prices by leveraging both external and internal market data. The UAMM algorithm calculates slippage while referencing fair prices as a benchmark. The detail of UAMM including the properties and guarantees can be found in the original paper \cite{uamm2023}. 

We describe UAMM for the binary outcome betting market, and yet, it extends naturally to the multi-outcome betting markets. 
Let us denote $\tau_{\text{In}}$ and $\tau_{\text{Out}}$ for input and output token index, where the $\text{Out}$ is the index of the event that the user is betting on. For example, if the user bets $1\tau$, then $1\tau_{\text{In}}$ and $1\tau_{\text{Out}}$ will be minted, $1\tau_{\text{In}}$ will be exchanged for some amount of $\tau_{\text{Out}}$ tokens, and returned $d\tau_{\text{Out}} = 1\tau_{\text{Out}} + \func{swap}(1\tau_{\text{In}}, \text{Out}; \Gamma)$ as shown in Algorithm~\ref{alg:odd_calc}. 
Here is a list of the set of parameters that represents the state of the market $\Gamma$:
\begin{itemize}
    \item $R\tau_{\text{In}}$ and $R\tau_{\text{Out}}$ be the liquidity pool balance of input and output tokens.
    \item $f\tau_{\text{In}}$ and $f\tau_{\text{Out}}$ be the probability of input and output event.
    \item $p\tau_{\text{In}}$ and $p\tau_{\text{Out}}$ be the spontaneous token prices w.r.t base currency $\tau_0$ given by the UAMM.
    \item $TS$ be the total supply of liquidity pool share tokens $s_{lp}$. 
    \item $TB$ be the total investment balance in terms of base currency $\tau_0$ (see Definition~\ref{def:tb}).
    \item $TV$ be the total value of the liquidity pool (see Definition~\ref{def:tv}).
\end{itemize}
Lastly, let us denote $d\tau_{\text{In}}$ as the input token amount for exchanging for $\tau_{\text{Out}}$ 
and denote $d\tau_{\text{Out}}$ be the output token amount that we exchanged from $\tau_{\text{In}}$.

\begin{algorithm}[t]
\caption{UBET AMM Buy Operation for Multiple Outcomes)}\label{alg:uamm_buy_op}
\begin{algorithmic}
\Require $i\in [1,K]$ is the outcome index to bet on.
\Require $d\tau_0 \geq 0$ is the wager size in base currency $\tau_0$.

\State Let $TS$ be total supply.
\State Let $TV$ be total value of the liquidity pool.
\State Let $\Gamma=\lbrace TS,TV,R\tau_0,R\tau_1,\cdots,R\tau_K\rbrace$ be UAMM parameters \\

\State //Mint $d$ amount of conditional tokens: $(d\tau_1, d\tau_2, \cdots, d\tau_K)$.
\State $(d\tau_1, d\tau_2, \cdots, d\tau_K) \leftarrow \func{mint}(d\tau_0)$
\State $odd = d\tau_i$\\
\State // Combine conditional token liquidity and collateral liquidity.
\State $R\tau_{k} \leftarrow R\tau_k + R\tau_0$ for all $k=[1,K]$\\
\State // Swap $\tau_j$ tokens for $\tau_i$ tokens 
\For{$j=[1,K]$, $j \neq i$}
    \State $R\tau_{j} \leftarrow R\tau_j + d\tau_0$
    \State $d\tau_i = \func{swap}(d\tau_j, \tau_i; \Gamma)$
    \State $odd \leftarrow odd + d\tau_i$.
\EndFor
\State $R\tau_{i} \leftarrow R\tau_i - d\tau_i$\\

\State // Merge any amount of conditional tokens from conditional token liquidity pool.
\State $(R\tau_0,R\tau_1,\cdots,R\tau_K) \leftarrow \func{merge}(R\tau_1, R\tau_2, \cdots, R\tau_K)$\\

\State // Update total value.
\State $TV \leftarrow R\tau_0 + \sum_k f\tau_k \cdot R\tau_k$\\
\State // Update total supply
\State $s_{lp} = d\tau_0 \cdot \frac{TS}{TV}$
\State $TS \leftarrow TS + s_{lp}$\\
\State Return ($odd$, $s_{lp}$)
\end{algorithmic}
\end{algorithm}

\subsection{Swap Function}
To remind you that swapping between the input and output tokens are necessary procedure in order to compute the odds, i.e., payout conditional token amounts. 
Despite relying on external market prices, our method maintains the desired properties of a constant product curve in computing slippages.
The swap function is defined as a piece-wise function such that
\begin{align*}
    \func{swap}(d\tau_{\text{In}}; \Gamma) =
    \begin{cases} 
        \alpha \cdot \Delta\tau_\text{Out} + (\rho -\alpha) (R\tau_\text{Out}-TB) & \text{ if } R\tau_{\text{Out}} - \Delta\tau_\text{Out}\leq TB \leq R\tau_{\text{Out}} \\
        \Delta\tau_{\text{Out}} & \text{ else if } TB \leq R\tau_{\text{Out}} \\
        R\tau_{\text{Out}} - \frac{TB^2}{X\tau_{\text{Out}}+\Delta\tau_{\text{Out}}}  & \text{ otherwise }
    \end{cases}
\end{align*}
where $\Delta\tau_{\text{Out}} = \rho \cdot d\tau_{\text{In}}$, $\rho = \frac{f\tau_\text{In}}{f\tau_\text{Out}}$ is the fair exchange rate for swapping $\tau_{\text{In}}$ for $\tau_{\text{Out}}$, $\alpha= \frac{R\tau_{\text{Out}}}{X\tau_{\text{Out}}+ \Delta\tau_{\text{Out}}}$, and $TB^2 = X\tau_{\text{Out}}\cdot R\tau_{\text{Out}}$ is a total investment balance of the pool.

We can understand the formulation of swap function $\func{swap}(d\tau_{\text{In}}, \text{Out}; \Gamma)$ as a two-step processes:
\begin{enumerate}
    \item compute the swapping amount based on price ratio, $\Delta\tau_{\text{Out}} = \rho \cdot d\tau_{\text{In}}$. 
    \item compute the slippage on output pool $R\tau_{\text{Out}}$, $d\tau_{\text{Out}} = \func{swap}(\Delta\tau_{\text{Out}}; \Gamma)$.
\end{enumerate}
Step 1 is the exchange rate times the wager size. 
Step 2 incorporates the slippage to the $\Delta\tau_{\text{Out}}$. 
Slippage refers to the difference between the expected odds of an asset and the actual odds at which a bet is executed. Slippage typically occurs in situations of high volatility, low liquidity, or when placing large orders.
UAMM adds slippage only then the liquidity pool balance of $\tau_{\text{Out}}$ is lower than the total investment balance $TB$ to make up the difference. Finally, we apply the swap function to $\tau_\text{In}$ and return the total $\tau_{\text{Out}}$ conditional token amounts to the bettors.

To elaborate further on matching the liquidity pool balance to the total investment balance, we must understand the concept of impermanent loss. Impermanent loss is a phenomenon that occurs when an LP deposits assets into a liquidity pool. It happens because of the fluctuation in the price of the assets being traded in the pool. When the price of one asset in the pool rises or falls more than the other, the LP's share of the pool becomes unbalanced. As a result, the LP's holdings of each asset are worth less than they would be if they had simply held them in their wallet. This reduction in the value of the LP's holdings is called impermanent loss. To minimize the impermanent loss, we want each liquidity pool balance to match the total investment balance. Due to the property that all probabilities/prices sum to 1, an equal balance of all conditional tokens can always be merged/redeemed to the same amount of collateral. When all pools have equal balances, no matter what the outcome probabilities are, the pool value is the same - there is no loss to liquidity providers if underlying probabilities change.

The full Algorithm of the buy operation is shown in Algorithm~\ref{alg:uamm_buy_op}.


\section{Experiment}
In this section, we look at the empirical LP impermanent performance of UAMM under various betting simulations and compare the UAMM. 
We demonstrate that UAMM maintains the total investment balance of LP funds and show UAMM incurs permanent gain while Uniswap incurs a permanent loss on average. 

\begin{figure}[t]
\centering
\begin{minipage}[c]{0.49\textwidth}
\centering
    \includegraphics[width=\textwidth]{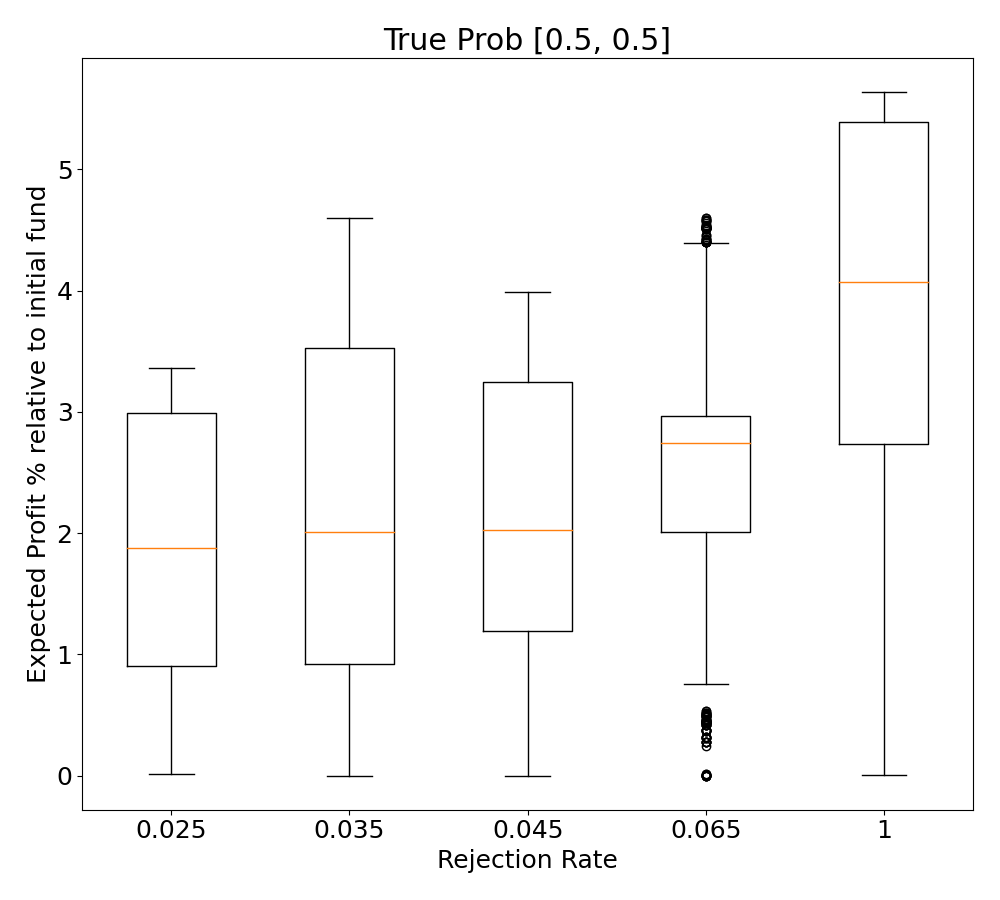}
    \subcaption{True Outcome Probability = (50\%, 50\%)}
    \label{fig:rejcetion_50_50}
\end{minipage}
\begin{minipage}[c]{0.49\textwidth}
\centering
    \includegraphics[width=\textwidth]{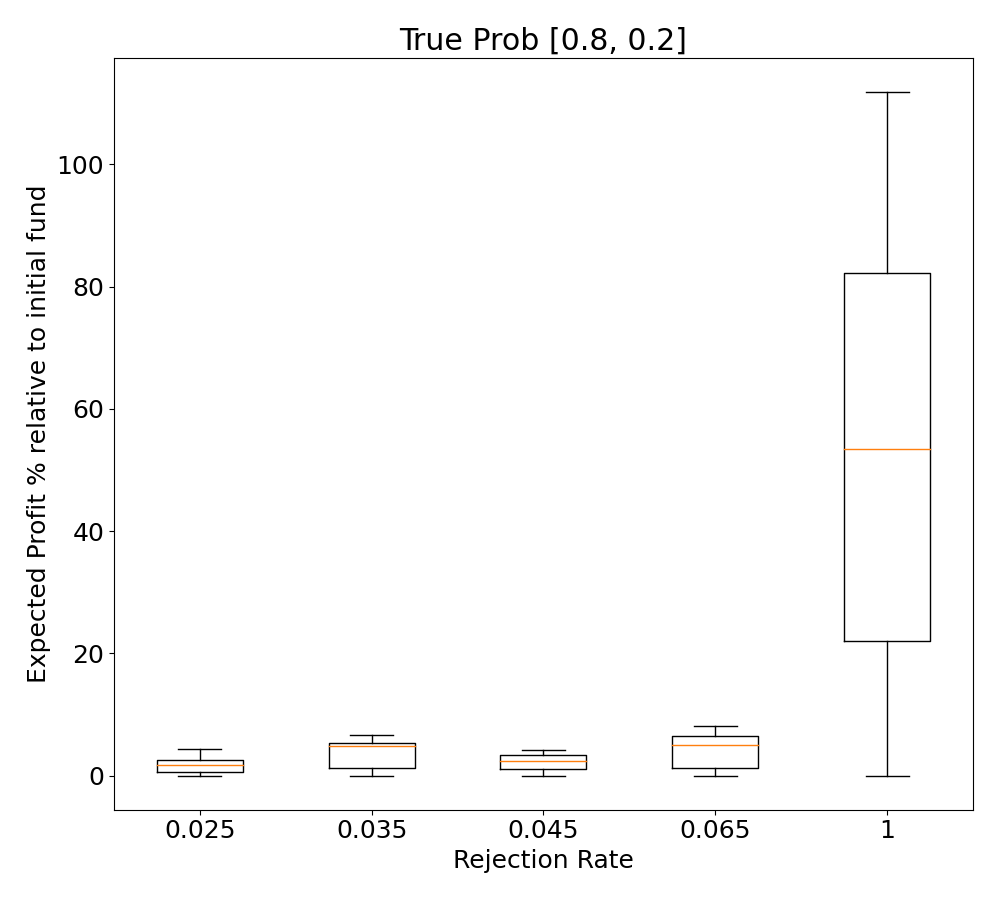}
    \subcaption{True Outcome Probability = (80\%, 20\%)}
    \label{fig:rejcetion_80_20}
\end{minipage}
\caption{Single run simulation of 500 betting transactions on two outcomes market w.r.t different odds rejection rate. The leftmost column displays the log ratio of the pool balance $log(x/y)-1$. The rightmost column displays EIP \% w.r.t to the LP funding amount.}
 \label{fig:rejection_two_outcomes}
\end{figure}

{\bf Experiment setup} 
Throughout the experiment, we assume that liquidity pools are funded initially when the market is created. During the experiments, there are multiple hyperparameters that we need to set for each market and betting simulation. These include the number of outcomes, the total funding, the actual probability of an outcome for a market, the number of total bets, the number of markets, the wager size distribution, and the bet side distribution.

We run our experiment over binary and ternary outcome betting markets. We consider funding amounts of $\$20,000$, $\$50,000$, and $\$100,000$ for each market. The true probabilities of outcome are chosen from a set of [20\%, 35\%, 50\%, 65\%, 80\%]. For the number of bets per market, we explore values of 10, 50, 100, and 500. We consider 10 bets because we may have fewer users at the beginning or some markets could be unpopular. Later, we re-run the experiments with the number of bets sampled from a scaled log-normal distribution with a mean of 2 and standard deviations of 1 which has approximately an average of 10 bets, a minimum of 1 bet, and a maximum of 40 bets.

We chose the number of markets to be [100, 500, 1000], assuming that the higher the number of markets, the average number of bets will be lower. To gather the empirical wager size distribution, we used data from the Polymarket sports prediction markets, where we scraped the transaction data from the Polygonscan. Simulating the bet side distribution is the most challenging part. The bettor may follow the probability of the outcome of events or could follow a 50-50 chance since the payout is already incorporated into the outcome selection. Thus, we simulate both cases where the bet sides are sampled from the true probability of outcome and from a 50-50 chance.

{\bf Metric} We observe the following metrics to evaluate UAMM:
\begin{itemize}
    \item Expected value of the pool:
    \begin{align*}
        \text{EV}(t) = \sum^{K}_{k}  \left(f\tau_{k,t} \cdot R\tau_{k,t} - (1-f\tau_{k,t}) \cdot (Z_t-R\tau_{k,t})\right)
    \end{align*}
    where $t \in [0, T]$ is the time, $T$ is the last time step when the betting market closes, and $Z_t=\sum^{K}_{k} R\tau_{k,t}$.
    \item Expected Impermanent Profit/Loss (PnL) with a standard deviation
    \begin{align*}
        \text{EIP}(T) = \frac{1}{M} \sum^{M}_{m}\sum^{K}_{k} f\tau_{k,T}^{(m)} \cdot (R\tau_{k,T}^{(m)}-R\tau_{k,0}^{(m)})
    \end{align*}
    where $M$ is the number of markets.
    \item Expected Permanent PnL with a standard deviation
    \begin{align*}
        \text{EPP}(T) = \frac{1}{M} \sum^{M}_{m} \sum^{K}_{k}  \mathbb{I}[s_T^{(m)} =\tau_{k}] \cdot (R\tau_{k,T}^{(m)}-R\tau_{k,0}^{(m)}) 
    \end{align*}
    where $s_T^{(m)}$ is the game result sampled from multinomial distribution $(f\tau_{1,T},\cdots,f\tau_{K,T})$.
    \item Total PnL is the actual gain or loss with the final game result: $\text{TP}(T) = M\cdot EPP(T)$. 

\end{itemize}
The difference between EIP and EPP is whether we use probability versus sampled binary outcome results to compute the PnL.
We understand EIP as the final game result has not come out, hence the PnL is impermanent while EPP calculates PnL based on the final result of the match.

First, we investigate the behaviour of a single betting market and then analyze the behaviour under multiple betting markets.

\begin{figure}[t]
\centering
\begin{minipage}[c]{\textwidth}
\centering
    \includegraphics[width=\textwidth]{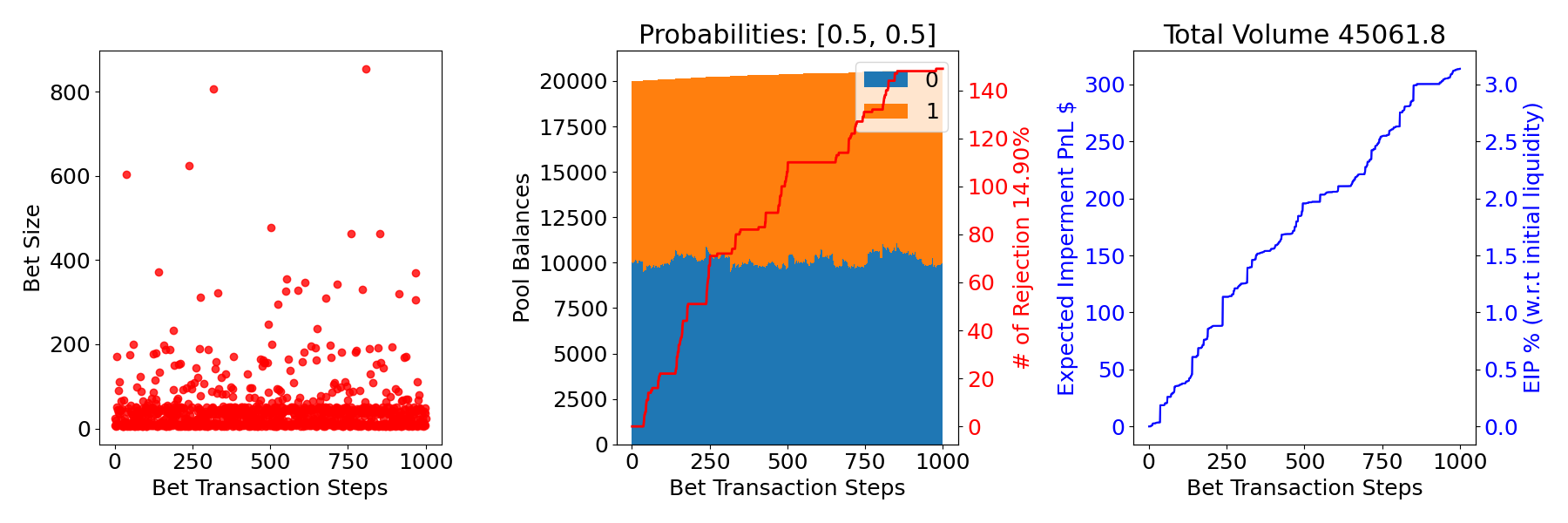}
    \subcaption{True Outcome Probability = (50\%, 50\%)}
    \label{fig:single_run_50_50}
\end{minipage}
\begin{minipage}[c]{\textwidth}
\centering
    \includegraphics[width=\textwidth]{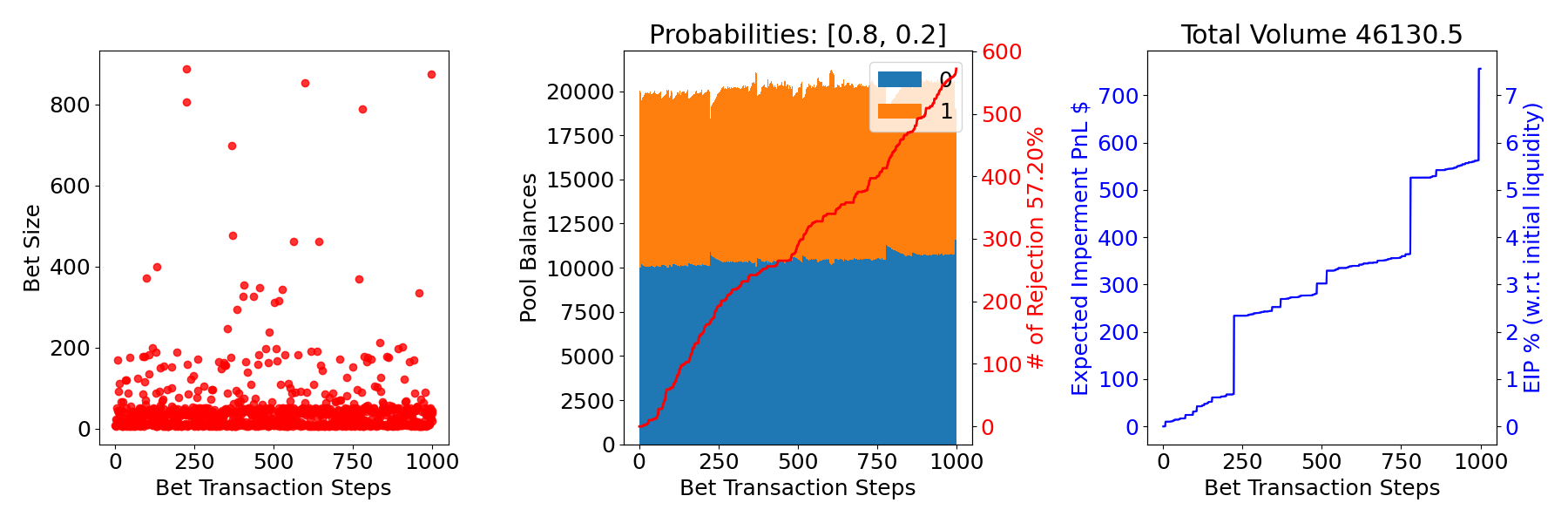}
    \subcaption{True Outcome Probability = (80\%, 20\%)}
    \label{fig:single_run_80_20}
\end{minipage}
\caption{Single run simulation of 1,000 betting transactions on two outcomes market. The leftmost column displays the sampled bet size distribution.
 The middle column displays the pool balance and the number of rejections over 100 transactions. The rightmost column displays EIP.}
 \label{fig:sing_run_two_outcomes}
\end{figure}

\subsection{Single Market Simulation}
For all our experiments, we assume that users will reject the odds when the spread is too high. To see the effect of applying odd rejection, we simulate 1,000 betting transactions with respect to different rejection rates for (50\%,50\%) and (80\%,20\%) outcome probabilities for two outcomes markets. Here, if the spread is larger than [0.025, 0.035, 0.045, 0.065, 1], then the bettor skips the bet. Figure~\ref{fig:rejection_two_outcomes} demonstrates that we make the most profit when the user doesn't reject any of our odds. We also see a trend where the profit increases as the acceptance rate increases. Throughout the experiments, we simulate this by sampling a rejection slippage from the Normal distribution with a mean of $0.045$ and a standard deviation of $0.05$. If the slippage is higher than the sampled value, the user rejects the bet.

Single-run simulation of 1,000 betting transactions on two outcomes market. Figure~\ref{fig:sing_run_two_outcomes} demonstrates two simulations with (50\%,50\%) and (80\%,20\%) outcome probabilities for two outcomes markets. The leftmost column displays the sampled bet size distribution. The middle column displays the pool balance and the number of rejections over 1,000 transactions. The rightmost column displays EIP. The liquidity pool balance was initialized to $\$10,000$, and we observe that the pool balances remain around the initial pool balance for both trials. The bet rejection rate for the (50\%, 50\%) market is approximately 14.9\%, while the bet rejection rate for the (80\%, 20\%) market is a lot higher, by 50\%. The total volume after 1,000 of the transactions came out to be $\$45,061$ and $\$46,130$, respectively.
 
Finally, the EIP came out to be positive. Additionally, Appendix Figure~\ref{fig:sing_run_three_outcomes} shows the same experiments for three outcomes markets. The results are similar to the two outcomes markets.

\begin{figure}[t]
\centering
\begin{minipage}[c]{\textwidth}
\centering
    \includegraphics[width=\textwidth]{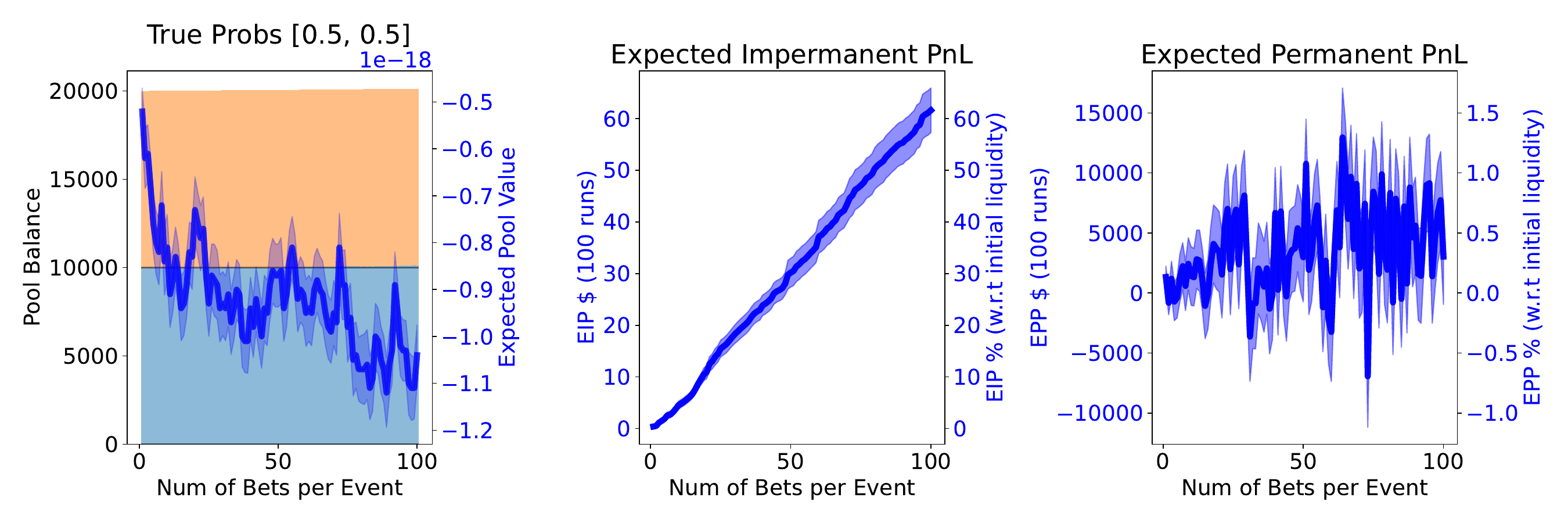}
    \subcaption{True Outcome Probability = (50\%, 50\%)}
    \label{fig:multiple_run_50_50}
\end{minipage}
\begin{minipage}[c]{\textwidth}
\centering
    \includegraphics[width=\textwidth]{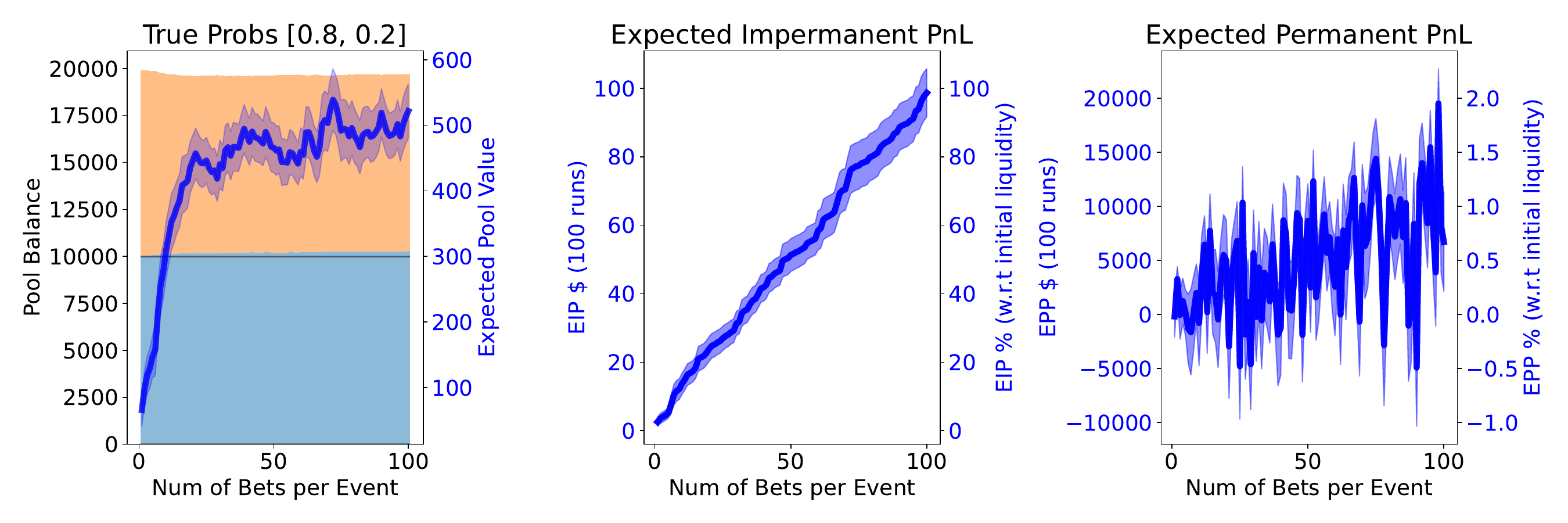}
    \subcaption{True Outcome Probability = (80\%, 20\%)}
    \label{fig:multiple_run_80_20}
\end{minipage}
\caption{Simulating multiple (100) betting markets with 100 betting transactions on two outcomes markets. 
 The leftmost column displays the pool balance and the expected value of the markets. 
 The middle and the rightmost columns show the expected impermanent PnL and expected permanent PnL.}
 \label{fig:multiple_run_two_outcomes}
\end{figure}

\subsection{Multiple Market Simulation}
The previous experiment was a single trial of 1,000 betting transactions. Here, we run the simulation over 100 trials. This is equivalent to running 100 betting markets, which consist of different game events and betting types and assuming that each market had 1,000 bets in total.
Figure~\ref{fig:multiple_run_two_outcomes} presents the results for (50\%, 50\%) and (80\%, 20\%) probability outcome events.
Due to central limit theorem \cite{Fendler08centrallimit}, the expected value of the multiple markets should remain zero and the average liquidity pool balance should remain $\$10,000$ initial liquidity pool for (50\%, 50\%) market. This is confirmed from the first column plot in Figure~\ref{fig:multiple_run_50_50} where the expected value is in the range of $1e-18$ and the ratio of the pool is well-balanced. 
We observe that the expected value is positive for the (80\%, 20\%) market (see Figure~\ref{fig:multiple_run_80_20}). This is because the pool balances are approximately even, we end up with a higher expected value when you multiply by the uneven probabilities.
It is good to observe that the EIPs are linearly going up as the number of bets increases.
Moreover, EPP for both markets is $50\%$ and $100\%$ of the initial liquidity pool.
The permanent PnL being highly volatile is expected since 100 coin flips with asymmetric bet size distribution are is have high variance.
We also show the results of running the same experiments for three outcome betting markets in Appendix Figure~\ref{fig:multiple_run_three_outcomes}.

\begin{figure}[t]
\centering
\begin{minipage}[c]{\textwidth}
\centering
    \includegraphics[width=\textwidth]{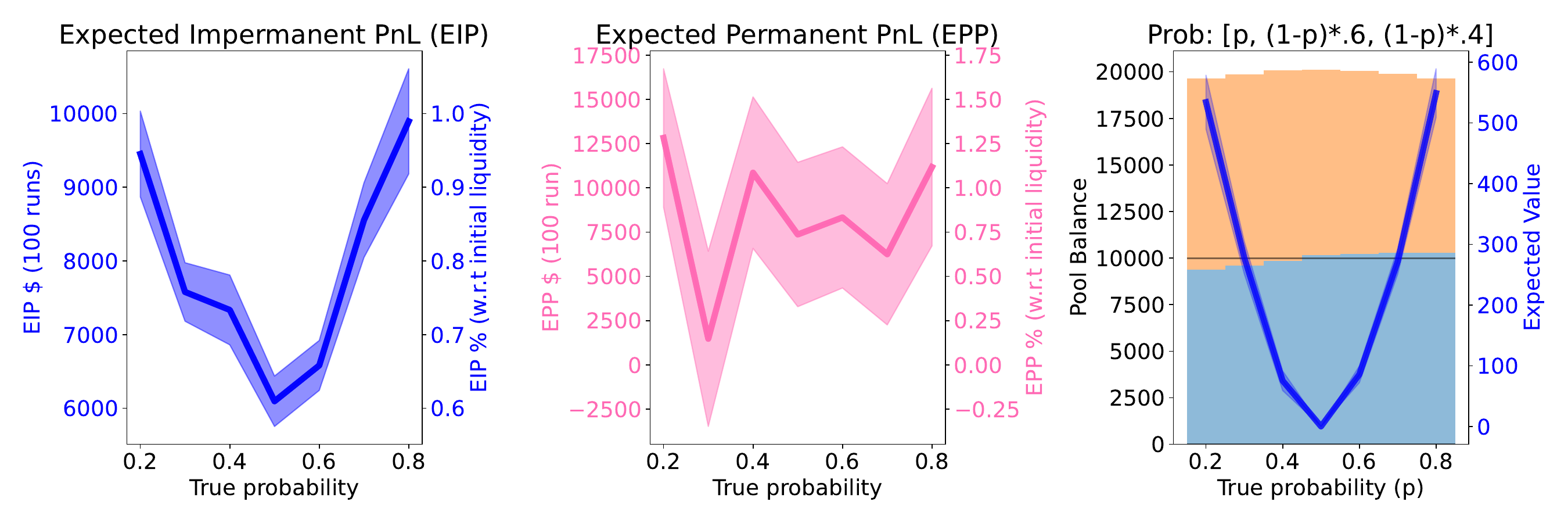}
    \subcaption{Bet side sampled from even distribution.}
    \label{fig:sample_even}
\end{minipage}
\begin{minipage}[c]{\textwidth}
\centering
    \includegraphics[width=\textwidth]{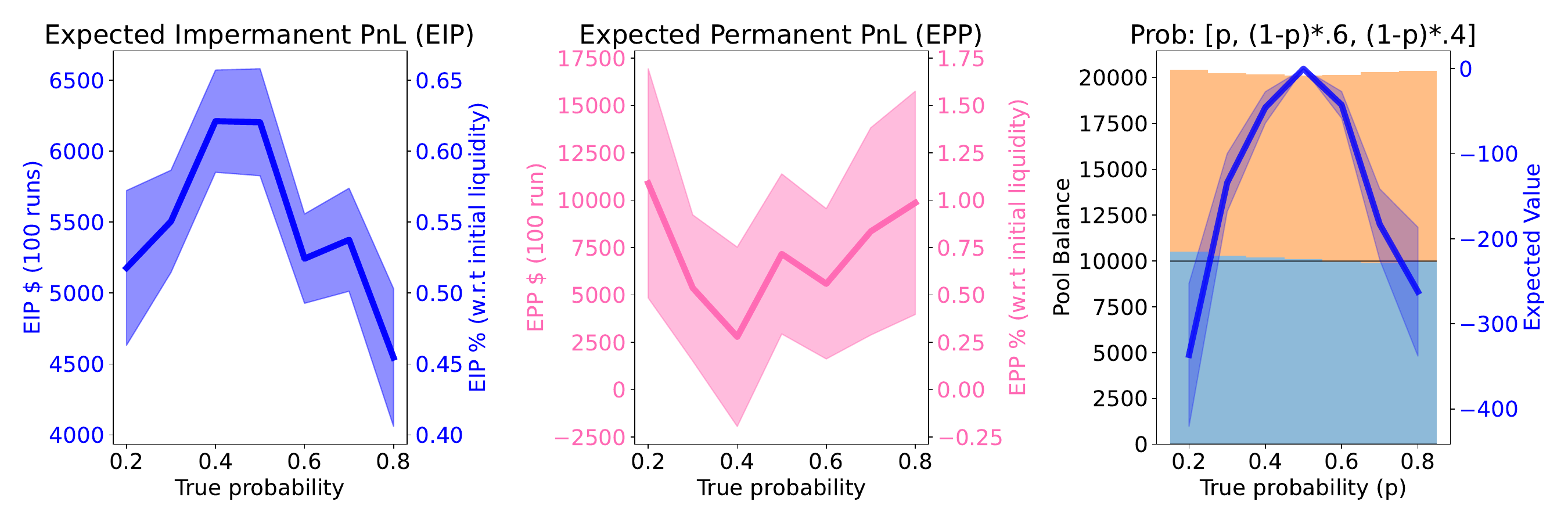}
    \subcaption{Bet side sampled from a true outcome probability distribution.}
    \label{fig:sample_true_prob}
\end{minipage}
\caption{The performance of multiple (100) betting markets simulation with 100 betting transactions on two outcomes markets with respect to various true outcome probabilities. 
 The leftmost and the middle columns show the expected impermanent PnL and expected permanent PnL.
 The rightmost column displays the pool balance and the expected value of the markets. }
 \label{fig:multiple_run_two_outcomes_wrt_probs}
\end{figure}

Next, we show sensitivity analysis with respect to different true outcome probability markets in Figure~\ref{fig:multiple_run_two_outcomes_wrt_probs}.
We conduct experiments over two bet side distributions, where bet sides are sampled from uniform vs true probability distribution.
Depending on the bettors, they maybe follow the probability of an event occurring versus others that do not care since the payout already incorporates the probabilities. 

Figure~\ref{fig:sample_true_prob} presents the EIP, EPP, EV, and pool balances over [0.2, 0.3, 0.4, 0.5, 0.6, 0.7, 0.8].
We observe that EIP and EV have quadratic curves where the higher PnL incurs with the skewed true probability events.
In contrast, EPP has a flat curve. Unfortunately, EIP and EV are actually artifacts and EPP presents the true nature of PnL. 
The artifact comes from the fact that we are multiplying uneven probabilities to even pool balances. However, in practice, the even tokens will get merged into collateral, which they do not have uncertainty associated with them.
In contrast to sampling from a uniform distribution, we observe that EIP and EV have reversed quadratic curves (see Figure~\ref{fig:multiple_run_two_outcomes_wrt_probs}). But again, the EPP has the same positive flat curve and see observe that
pool balances are even. Therefore the negative EV can be ignored.
Appendix Figure~\ref{fig:multiple_run_three_outcomes_wrt_probs} has the results for three outcome markets.

\subsection{Full market simulation}
So far, we executed controlled experiments except for the bet sizes and showed the UAMM's ideal behaviours. 
Now we run an uncontrolled experiment where all hyperparameters are sampled from the following distributions:
the number of outcomes is sampled randomly between 2 and 3,
the true outcome probabilities for betting markets are sampled uniformly between (0.2,0.8), and
the number of bet transactions is sampled from a log-normal distribution with a mean of two and a standard deviation of two.

\begin{table}[ht]
\centering
\begin{tabular}{||c | c | c | c | c||} 
 \hline
 Method & \# of Bets & Volume & EPP \$ (\%) & EPP + Fee \$ (\%) \\ [0.5ex] 
 \hline
 UAMM & 1203 $\pm$ 15 & \$54107 $\pm$ 742.2 & \$19.93 $\pm$ 3.13 (0.0019\%) & 1372 $\pm$ 18.8 (0.137\%)\\ [1ex] 
 \hline
\end{tabular}
\vspace{0.3cm}
\caption{ Performance over running 100 uncontrolled trials with $10,000$ initial liquidity funding. }
\label{tab:performance}
\end{table}

Table~\ref{tab:performance} presents the results of running market simulations 100 times.
On average, each market has about an average of 10 bets, and minimum and maximum of 2 and 40 bets.
After running 100 trials with 100 betting markets, we observe 1203 bet transactions in total and \$54,107 total volumes,
and \$20 gains in average, and \$1,372 gains including the 2.5\% transaction fees.

\section{Conclusion}
The sports betting market's rapid growth and widespread adoption of online platforms and mobile apps have propelled its expansion. Nevertheless, traditional sports betting faces significant challenges, including centralization, lack of transparency, high fees, and regulatory constraints.

Decentralization emerges as a transformative solution to address these limitations in sports betting. By leveraging blockchain technology and smart contracts, decentralization ensures transparency, security, and fairness through immutable records of all transactions and bets. The elimination of intermediaries and lower fees make decentralized platforms more inclusive and appealing to users worldwide. Moreover, non-custodial solutions empower bettors to have greater control over their funds.

AMMs play a pivotal role in the decentralized sports betting ecosystem, providing efficient and continuous liquidity provision. Our proposed solution, UAMM, addresses the limitations of existing AMMs in pricing sports betting odds. By utilizing UAMM algorithms and referencing other sportsbook odds, we achieve fair odds and minimize risks for liquidity providers.

In this paper, we introduced a formal methodology for creating sports betting markets on a blockchain, demonstrated the novel application of UAMM for sports betting, and proposed an additional collateral liquidity pool to streamline the betting process. Through extensive experiments and analysis, we showcased positive permanent gains while maintaining a relatively low vigorish.

Overall, decentralized sports betting, powered by UAMMs and smart contracts, holds the potential to revolutionize the industry. It provides users with a trustless, transparent, and efficient betting environment, empowering them to engage in a global and democratic betting ecosystem. As decentralization continues to shape the future of sports betting, we anticipate continued growth and innovation, driving the industry toward greater accessibility, fairness, and user satisfaction.

\bibliographystyle{plain}
\bibliography{main}

\appendix
\section{Appendix}

\subsection{Notations of things}
We define metrics that track the states of LPs and liquidity pools. 

\begin{definition}[Total Investment Balance a.k.a, Target Balance] \label{def:tb}
    Let $A_{\ttime{1}}, A_{\ttime{2}}, \cdots, A_{\ttime{T}}$ be the sequence of actions.
    The total investment balance at the time $\ttime{T}$,
    \begin{align*}
        TB_{\ttime{T}} := \sum^T_{\ttime{t}=1} \left[ \Big(\mathbb{I}[A_{\ttime{t}}=\func{Add}] - \mathbb{I}[A_{\ttime{t}}=\func{Remove}]\Big) \sum^K_{i=1} d\tau_{i,\ttime{t}} \cdot f\tau_{i,\ttime{t}} \right]
    \end{align*}
    where $f\tau_{i,\ttime{t}}$ is the fair price and $R\tau_{i,\ttime{t}}$ is the pool balance of $\tau_i$ at time $\ttime{t}$.
\end{definition}

\begin{definition}[Liquidity Pool Value] \label{def:tv}
    The total value of the liquidity pool is 
    \begin{align*}
        TV := \sum^K_{i=1} f\tau_{i} \cdot R\tau_{i}
    \end{align*}
    where $f\tau_i$ is the fair price and $R_i$ is the pool balance of $\tau_i$.
    If the fair prices are probability\footnote{the total fair prices sum up to 1 and each fair price lies in between $[0,1]$}, then the total value is the expected value (EV) of the liquidity pool.
\end{definition}

\subsubsection{Experiment Supplementary Materials}
\begin{figure}[ht]
 \begin{minipage}[c]{\textwidth}
\centering
    \includegraphics[width=\textwidth]{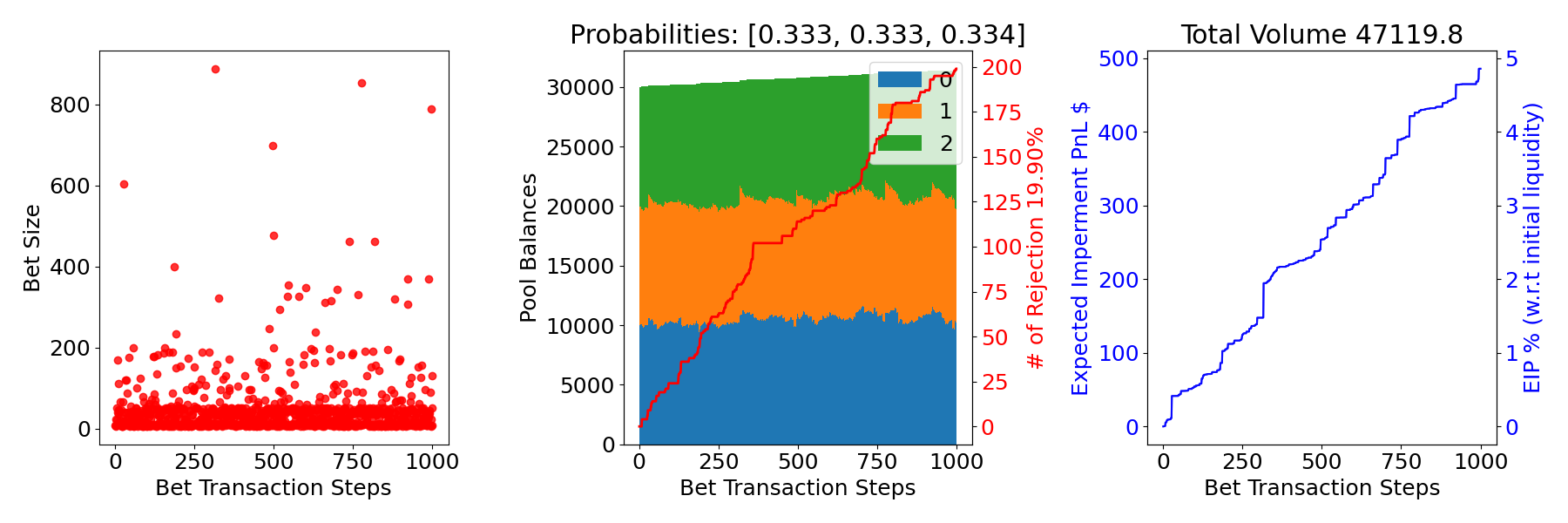}
    \subcaption{True Outcome Probability = (33\%, 33\%, 33\%)}
    \label{fig:single_run_33_33_33}
\end{minipage}
\begin{minipage}[c]{\textwidth}
\centering
    \includegraphics[width=\textwidth]{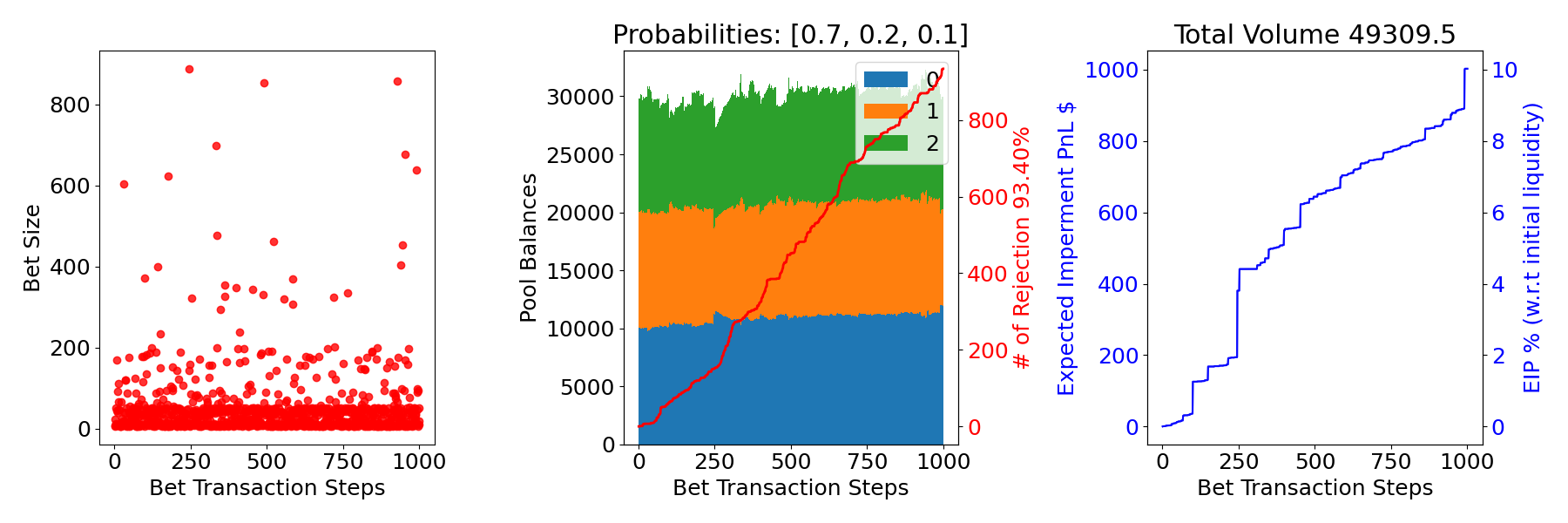}
    \subcaption{True Outcome Probability = (70\%, 20\%, 10\%)}
    \label{fig:single_run_70_20_10}
\end{minipage}
\caption{Single run simulation of 1,000 betting transactions on three outcomes market.}
\label{fig:sing_run_three_outcomes}
\end{figure}

\begin{figure}[t]
\centering
\begin{minipage}[c]{\textwidth}
\centering
    \includegraphics[width=\textwidth]{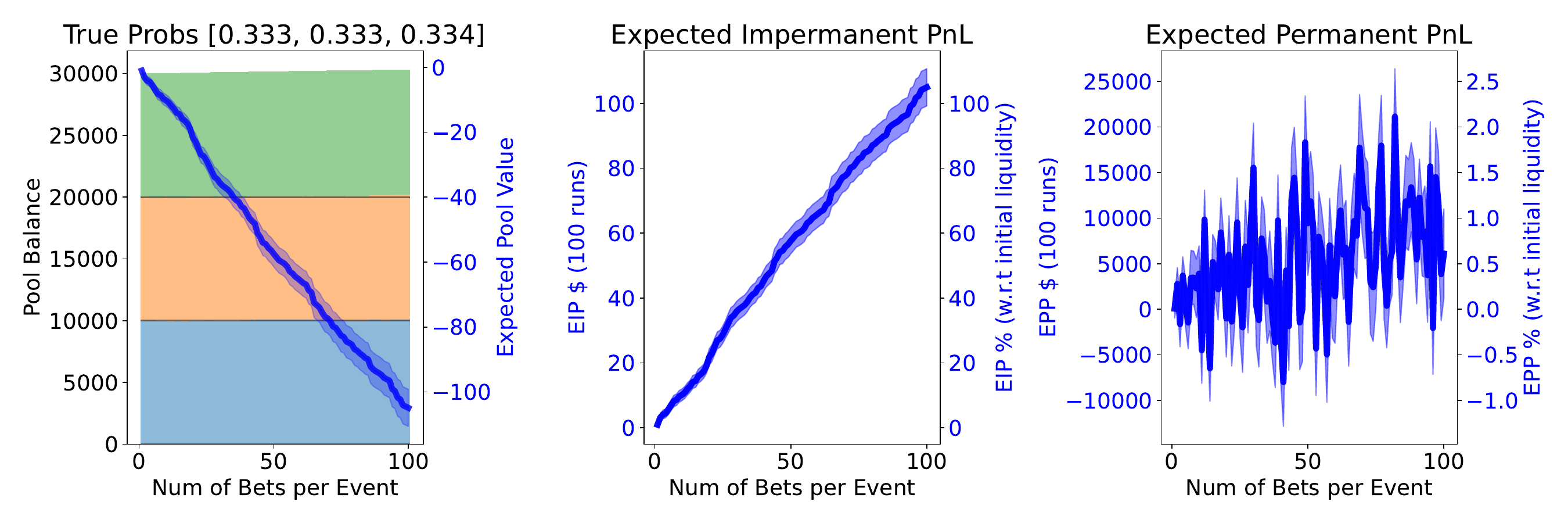}
    \subcaption{True Outcome Probability =  (33\%, 33\%, 33\%)}
    \label{fig:multiple_run_33_33_34}
\end{minipage}
\begin{minipage}[c]{\textwidth}
\centering
    \includegraphics[width=\textwidth]{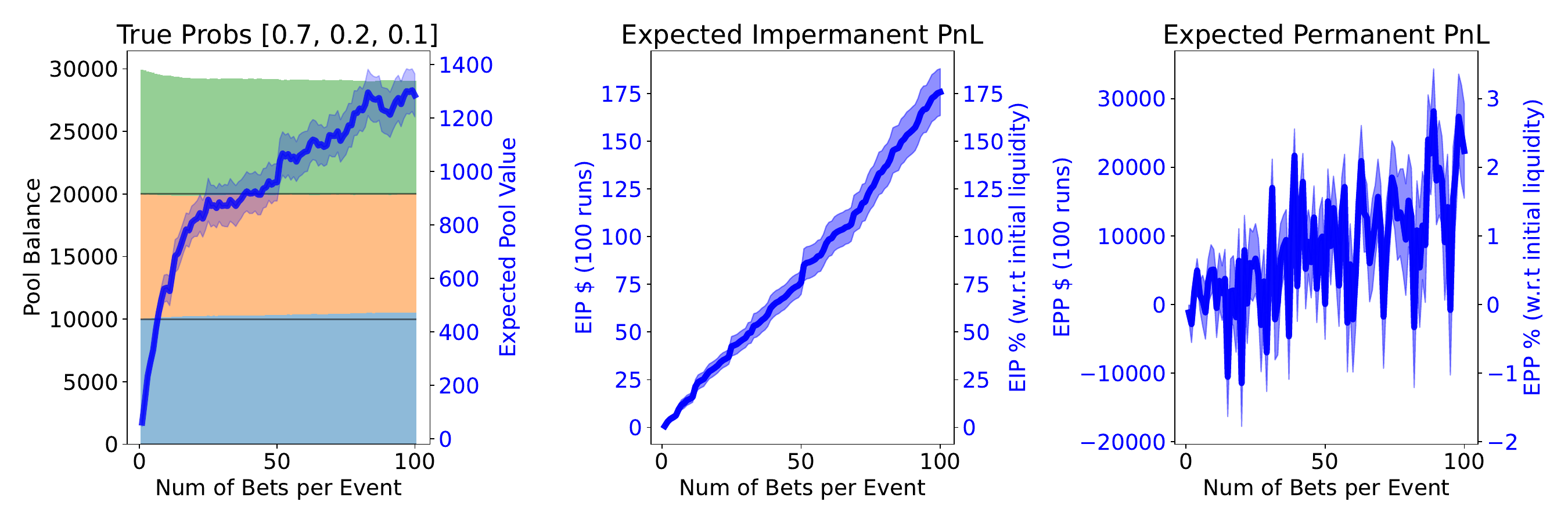}
    \subcaption{True Outcome Probability = (70\%, 20\%, 10\%)}
    \label{fig:multiple_run_70_20_10}
\end{minipage}
\caption{Simulating multiple (100) betting markets with 100 betting transactions on three outcomes markets. 
 The leftmost column displays the pool balance and the expected value of the markets. 
 The middle and the rightmost columns show the expected impermanent PnL and expected permanent PnL.}
 \label{fig:multiple_run_three_outcomes}
\end{figure}

\begin{figure}[t]
\centering
\begin{minipage}[c]{\textwidth}
\centering
    \includegraphics[width=\textwidth]{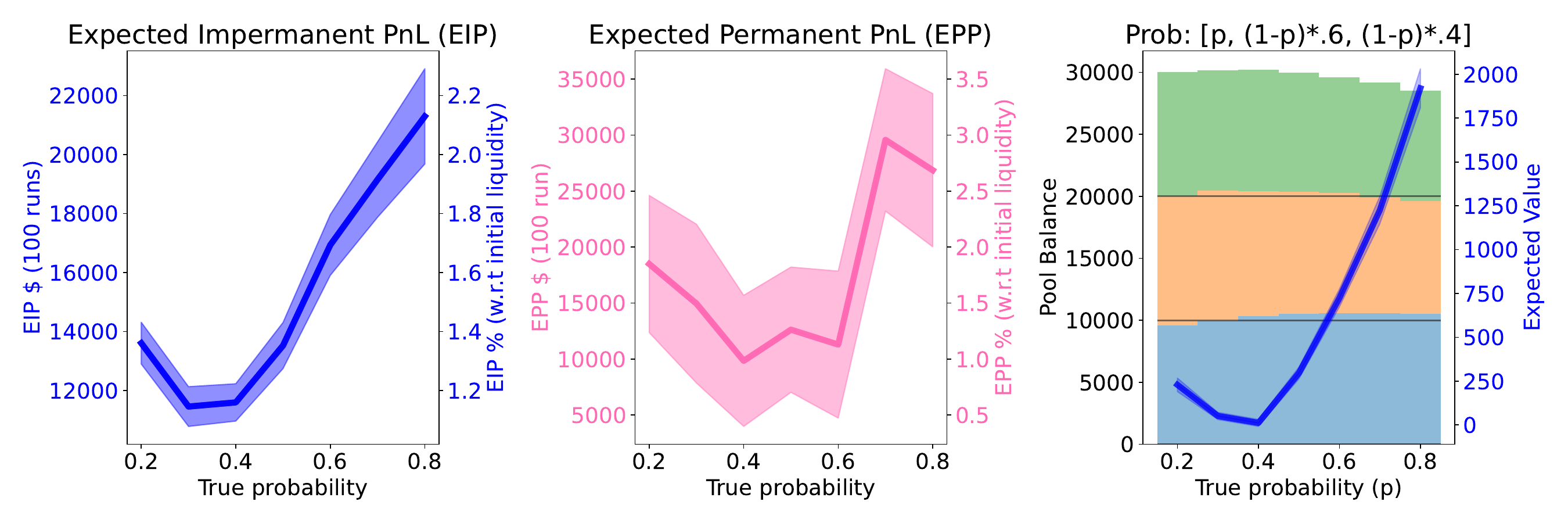}
    \subcaption{Bet side sampled from even distribution.}
    \label{fig:sample_even3}
\end{minipage}
\begin{minipage}[c]{\textwidth}
\centering
    \includegraphics[width=\textwidth]{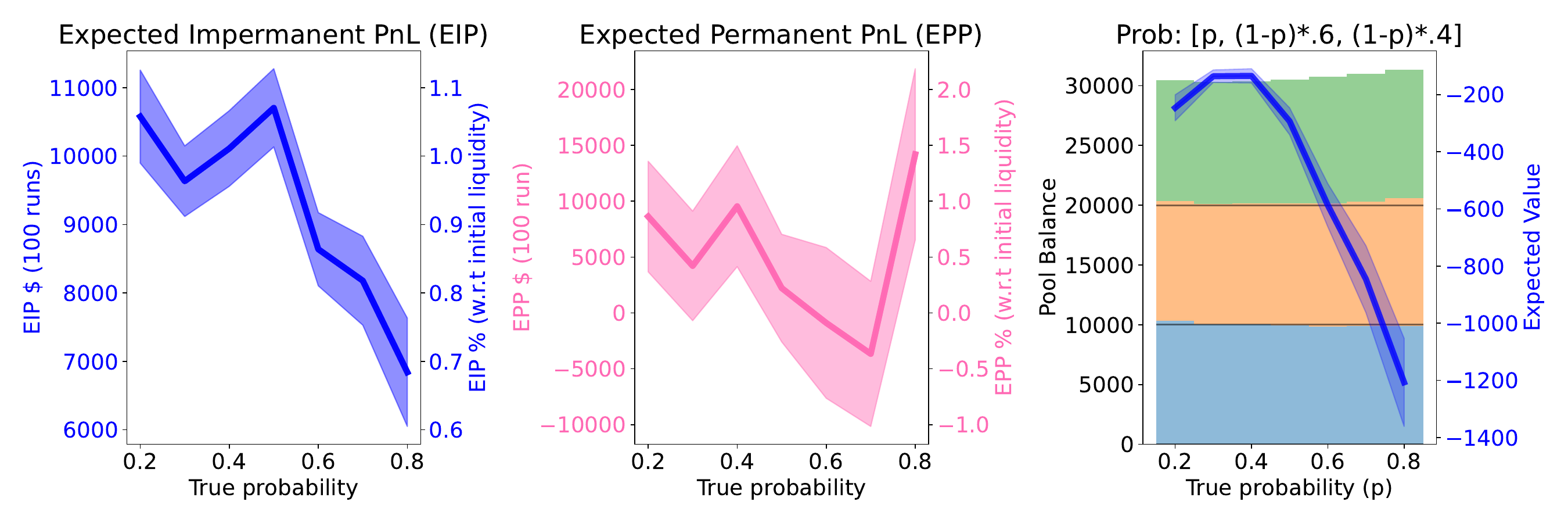}
    \subcaption{Bet side sampled from a true outcome probability distribution.}
    \label{fig:sample_true_prob3}
\end{minipage}
\caption{The performance of multiple (100) betting markets simulation with 100 betting transactions on three outcomes markets with respect to various true outcome probabilities. 
 The leftmost and the middle columns show the expected impermanent PnL and expected permanent PnL.
 The rightmost column displays the pool balance and the expected value of the markets. }
 \label{fig:multiple_run_three_outcomes_wrt_probs}
\end{figure}


\end{document}